\documentstyle[12pt]{article}
  \advance\oddsidemargin  by -1in
  \advance\evensidemargin by -1in
\def\lsim{\mathrel{\rlap{\lower4pt\hbox{\hskip1pt$\sim$}}
    \raise1pt\hbox{$<$}}}         
\def\gsim{\mathrel{\rlap{\lower4pt\hbox{\hskip1pt$\sim$}}
    \raise1pt\hbox{$>$}}}         

\def\be{\begin{equation}}
\def\ee{\end{equation}}
\def\bq{\begin{eqnarray}}
\def\eq{\end{eqnarray}}

\mathchardef\alpha="710B
\mathchardef\beta="710C
\mathchardef\gamma="710D
\mathchardef\delta="710E
\mathchardef\epsilon="710F
\mathchardef\zeta="7110
\mathchardef\eta="7111
\mathchardef\theta="7112
\mathchardef\iota="7113
\mathchardef\kappa="7114
\mathchardef\lambda="7115
\mathchardef\mu="7116
\mathchardef\nu="7117
\mathchardef\xi="7118
\mathchardef\pi="7119
\mathchardef\rho="711A
\mathchardef\sigma="711B
\mathchardef\tau="711C
\mathchardef\upsilon="711D
\mathchardef\phi="711E
\mathchardef\chi="711F
\mathchardef\psi="7120
\mathchardef\omega="7121
\mathchardef\varepsilon="7122
\mathchardef\vartheta="7123
\mathchardef\varpi="7124
\mathchardef\varrho="7125
\mathchardef\varsigma="7126
\mathchardef\varphi="7127
\mathchardef\nabla="7272

\font\dozeb=cmmib10 scaled \magstep1
\font\dozesyb=cmbsy10 scaled \magstep1
\font\dezb=cmmib10
\begin{document}
\pagestyle{empty}
\hfill{\bf DFTT 35/95}

\hfill{{\bf CERN-TH/95-164}

\vspace{2.0cm}

\begin{center}

{\large \boldmath \bf EXTRACTING THE STRANGE DENSITY FROM $xF_3$ \\}

\vspace{1.0cm}

{\large V.~Barone$^{a,}$ \footnote{Also at
II Facolt{\`a} di Scienze MFN, I--15100 Alessandria, Italy.},
U.~D'Alesio$^{a}$ and M.~Genovese$^{a,b}$\\}

\vspace{1.0cm}

{\it $^{a}$Dipartimento di
Fisica Teorica dell' Universit\`a \\
and INFN, Sezione di
Torino,   I--10125 Turin, Italy \medskip\\
$^{b}$Theory Division, CERN, CH--1211 Geneva
23, Switzerland \medskip\\ }

\vspace{1.0cm}

{\large \bf Abstract \bigskip\\ }
\end{center}

We present a  QCD analysis
of the strange and charm contributions to the neutrino
deep
inelastic structure function $xF_3$.
We show that next-to-leading order effects, which are
relatively important for $F_2$, play a lesser
role in the case of $xF_3$. The neutrino--antineutrino
difference
$xF_3^{\nu} - xF_3^{\bar \nu}$ provides a
new determination
of the strange density, which exhibits some advantages
with respect to other traditional
methods.

\vskip 3cm
{\bf CERN-TH/95-164}

{\bf June 1995}

\vfill

\pagebreak

\baselineskip 24 pt
\pagestyle{plain}
Two methods are traditionally used to extract the strange-sea density
from deep inelastic scattering (DIS) data:
the first
consists in studying charm production in charged-current neutrino
DIS (the characteristic signature of this process being the
presence of dimuons in the final state);
the second
is to subtract
the $F_2$ structure functions
measured in neutrino and muon DIS, thus
selecting the strange
contribution.

Until last year, these two determinations, based on the
NMC $\mu$DIS
data \cite{NMC} and on
the CCFR $\nu$DIS data \cite{CCFR1,CCFR3} available at that time,
 seemed to yield contradictory results for
$s(x)$. This discrepancy, which has strongly challenged
 the attempts at
global parton fits \cite{MRS1,CTEQ}, had been actually predicted
\cite{BGNPZ1,BGNPZ2}, and is simply
explained \cite{BGNPZ3,BGNPZ4}
by the observation that the dimuon and $\nu-\mu$ determinations
of $s(x)$ actually measure different quantities,
related to -- but
not coincident with -- the strange density. This is due to the
relevance of
quark-mass effects \footnote{Remember that in
charged-current neutrino DIS strange and charm excitations
are inseparable.} and longitudinal contributions,
in the region of small and moderate $Q^2$ values
(of order of $10$-$30$  GeV$^2$)
\footnote{For instance, the average $Q^2$ value
of the CCFR data is
$\langle Q^2 \rangle \simeq 22 \, $GeV$^2$.}.

A recent QCD analysis \cite{BGNPZ5} of the new CCFR dimuon data
\cite{CCFR2}
 has shown that, when all important physical effects
are taken into account,
the different
measurements converge -- as they should -- towards a unique
result for $s(x)$. Incidentally, the strange-quark distribution
emerging from all data
is well reproduced by
a ``traditional'' fit, such as, for instance, MRS(A) \cite{MRS2},
and does not support a nearly
$SU(3)$ symmetric fit, such as CTEQ1 \cite{CTEQ}.

Both the dimuon and the $\nu$-$\mu$ extractions of $s(x)$
present problems and subtleties (for a detailed discussion
see \cite{BGNPZ4,BGNPZ5}). In particular, the dimuon determination
implies, for experimental reasons, an acceptance-dependent
separation of the {\it t}- and {\it u}-channel diagrams that constitute
the $W$--gluon fusion QCD process.
On the other hand, the $\nu-\mu$ result
is affected by large uncertainties due to the very unsafe
procedure of subtracting data from two different experiments.

However, the idea of obtaining $s(x)$ by an appropriate combination
of structure functions can be further exploited. There is in fact
another way to isolate $s(x)$ from DIS structure functions, which
makes use of the third $\nu$DIS structure function,
$F_3$. In the parton model these are \cite{libri}
\bq
xF_3^{\nu N}(x)&=& xV(x) - 2 x \bar c(x) + 2 x s(x)\,,
\label{1} \\
x F_3^{\bar \nu N}(x) &=& x V(x) + 2 x c(x) - 2 x \bar s(x)\,,
\label{2}
\eq
where $V(x)$ is the
valence distribution and
$N$ denotes an isoscalar nucleon (hereafter we shall drop
the suffix $N$ from our formulas). Therefore,
the $\nu - \bar \nu$ difference
effectively measures the strange density, since the charm contribution
is very small, at least
in the kinematical region investigated by the present  experiments
(we assume $s = \bar s$ and $c = \bar c$):
\be
xF_3^{\nu N}(x) - x F_3^{\bar \nu N}(x) = 4 x \left [ s(x) -
c(x) \right ]\,.
\label{3}
\ee
Needless to say, moving from the parton model to
leading-order QCD,
the quark distributions
acquire a  $Q^2$ dependence governed by the Altarelli--Parisi
equations.

The use of eq.~(\ref{3}) to extract $xs(x)$ has the immediate advantage,
over the $\nu-\mu$ method, of combining data from the same
experiment, thus with no relative-normalization problems. However,
in practice there are at least
two shortcomings. First of all, $xF_3$ is a small,
valence-dominated, quantity,
more difficult to measure than $F_2$. Secondly, the statistics for
$xF_3^{\nu}- xF_3^{\bar  \nu}$ is
much lower than that
for $F_2$: the measurement of $xF_3^{\nu} -
xF_3^{\bar \nu}$ requires neutrino and antineutrino data separately,
whereas they can be combined for $F_2$, which is the same
in $\nu N$ and $\bar \nu N$ deep inelastic scattering. In general,
the number of $\bar \nu$-induced events is considerably
smaller than that of $\nu$-induced events: for instance,
the ratio is about 1 to 5 in the CCFR experiment \cite{CCFR2}.

The CCFR/NuTeV Collaboration is at present working on the extraction
of $xs(x)$ from $xF_3^{\nu N} - xF_3^{\bar \nu N}$, and data will
be available in the near future \cite{bazarko}.

Previous studies on the determination
of the strange density at moderate $Q^2$ from neutrino DIS
have taught us the
importance of quark-mass corrections and current non-conservation
effects, which manifest themselves through the
order-$\alpha_s$ vector-boson--gluon fusion diagrams.
These represent the dominant contribution
near the heavy-quark threshold, in particular at small $x$.
It is then natural to go beyond the leading order also
in the analysis of the extraction of $xs(x)$ from $xF_3$.
The Next--to--Leading Order corrections are known
to affect
mostly the longitudinal component of structure functions
\cite{BGNPZ2}: thus we expect the
NLO effects
to be smaller in $xF_3$ than in $F_2$,
because
$xF_3$ is a purely transverse structure function.
However, only an explicit computation
can give us precise information on the
charm--strange content of $xF_3$.

In the following we shall present a QCD calculation of
$xF_3^{\nu}-xF_3^{\bar \nu}$ at order
$\alpha_s$, taking into account the contribution of
the gluon-fusion diagrams. The possibility of
a safe, unambiguous, extraction of the strange density from
$xF_3^{\nu N} - x F_3^{\bar \nu N}$ will be explored and
discussed.

At order $\alpha_s$ the main contribution to the $cs$ component
of $F_3^{\nu}$ is
given by the $W$--gluon fusion (GF) term \footnote{The
$O(\alpha_s)$ $W$--quark
fusion diagrams are a negligible correction.}, which,
for the strange--charm sector, reads
\be
F_{3,GF}^{\nu,cs}(x,Q^2) =
\left(\frac{\alpha_s}{2 \pi} \right) \,
\int_{ax}^{1}
\frac{{\rm d}z}{z} \, g (z, \mu^2) \,
C_3 \left(\frac{x}{z}, Q^2 \right)\,.
\label{a0}
\ee
Here $a = 1 + (m^2 + m'^2)/Q^2$ and,
 for neutrino scattering, $m$ is the mass of the charmed quark,
$m'$ is the mass of the strange antiquark.
The Wilson coefficient $C_3$ represents the
$W^{+}g \rightarrow c \bar s$
cross section difference $\sigma_L
- \sigma_R$ ($L$ and $R$ standing
for left- and right-transverse $W$, respectively); it
has the explicit
form \cite{WGR,GR}:
\bq
C_3(z,Q^2) &=& 2 \, \left \{
\beta_{+}  \beta_{-} \, \frac{m^2-m'^2}{Q^2} \,
2 z (1-z) \right.  \nonumber \\
&-& \left. \left [ \frac{1}{2} - z (1-z) + \frac{m^2 -
m'^2}{Q^2} \, z (1-2z) - \frac{m^4 - m'^4}{Q^4} \, z^2
\right ]  \, L(m,m') \right. \nonumber \\
&+&  \left. \left [ m \leftrightarrow m' \right ]
\, L(m',m) \right \} \,,
\label{a1}
\eq
where
\be
\beta_{\pm}^2 = 1 - \frac{(m \pm m')^2}{Q^2} \, \frac{z}{1-z}
\label{a2}
\ee
and
\be
L(m,m') = \log {\frac{1+ \frac{m^2 - m'^2}{Q^2} \, \frac{z}{1-z}
+ \beta_{+} \beta_{-} }{1+ \frac{m^2 - m'^2}{Q^2} \, \frac{z}{1-z}
- \beta_{+} \beta_{-} }}\,.
\label{a3}
\ee

If the two quark masses are non-zero, the
Wilson coefficient $C_3$ is free from singularities.
In the limit $m' \rightarrow 0$, {\it i.e.} treating the
strange quark as massless,
$L(m,m')$ and $L(m',m)$ behave as
\bq
L(m,m') &\rightarrow &
\log { \frac{\hat s}{m^2}}
\label{a4} \\
L(m',m) &\rightarrow &
\log{ \frac{(\hat s - m^2)^2}{\hat s \, m'^2} }
\label{a5}
\eq
where $ \hat s = Q^2 (1-z)/z$.
In the same limit, the factor multiplying $L(m',m)$ becomes
${\cal P}_g(\xi)$, namely
the usual $g \rightarrow q \bar q$ splitting function ${\cal P}_g$
expressed in the rescaled variable
$\xi = x (1 + m^2/Q^2)$.
While $L(m,m')$
is regular, $L(m',m)$ has a collinear singularity. This
is subtracted out by setting \cite{Tung}
\be
L(m',m) = \log { \frac{(\hat s - m^2)^2}{\hat s \, \mu^2} }\,,
\label{a6}
\ee
where the scale $\mu^2$
is customarily taken to be equal to the factorization scale
({\it i.e.} to the scale that separates the perturbative part
from the non-perturbative one in the DIS QCD diagrams).

Let us now come to the total $cs$ contribution to $F_3$.
We start from the
QCD factorization formula, which formally reads
($\otimes$ means convolution)
\be
F(x,Q^2) = \sum_i f_i(z,\mu^2) \otimes C(\frac{x}{z}, \mu^2, Q^2)\,,
\label{a6b}
\ee
where the sum is made over all parton species.

At leading order only quark and antiquark contribute and the
corresponding Wilson coefficients are delta functions of
the slow-rescaling variable $\xi = x (1 + m^2/Q^2)$.
At next-to-leading order the contribution is given by eq.~(\ref{a0})
with the above-defined subtraction.
Thus we get
\bq
F_3^{\nu,cs}(x,Q^2) &\equiv& F_{3,QE}^{\nu,cs}(x,\mu^2)
+ \tilde F_{3,GF}^{\nu,cs}(x,Q^2) \nonumber \\
&=& 2 \left [ \bar s(\xi, \mu^2) - c (x, \mu^2)
\right ] + \left(\frac{\alpha_s}{\pi} \right) \,
\int_{a x}^{1} \frac{{\rm d}z}{z} \, g (z, \mu^2) \,
\tilde C_3 \left(\frac{x}{z}, Q^2 \right)\,.
\label{a7}
\eq
Here $\tilde C_3$ stands for the subtracted Wilson coefficient,
{\it i.e.} for $C_3$ with
the replacement (\ref{a6}).
Notice that the quark excitation (QE) term, $\bar s - c$, is taken at
the factorization scale $\mu^2$. The $Q^2$ evolution
(at least the dominant part of it, due to $g \rightarrow q \bar q$
splitting) is already contained in the gluon-fusion Wilson coefficient.
Both taking the quark densities in eq.~(\ref{a7}) at the
physical scale $Q^2$ and using $C_3$ instead of $\tilde C_3$ in
eq.~(\ref{a7})
  would represent a double
counting. A most often used approximation consists in
setting
\be
F_3^{\nu,cs}(x,Q^2) = 2 \left [ \bar s(\xi,Q^2) - c(x,Q^2)
\right ] \,,
\label{a8}
\ee
which means combining massless QCD, for the
$Q^2$ evolution, with slow rescaling, to account for quark mass
effects. In the following
we shall check the
goodness of this slow rescaling procedure for $F_3$.

At large momentum transfer ($Q^2 \gg m_c^2$, say $Q^2 \gsim
30$ GeV$^2$) the charmed quark should also be treated as
a massless parton. This means that $L(m,m')$ too becomes large and
should undergo a subtraction similar to that performed by
eq.~(\ref{a6}).
Asymptotically, the massless QCD formulas are of course regained:
in the limit $m, m' \rightarrow 0$ the contribution of the
gluon-fusion diagram to $F_3$ vanishes, {\it i.e.}
$\tilde F_{3,GF}^{\nu,cs} \rightarrow 0$.

So much for the neutrino DIS. In the case of antineutrino scattering
the formulas written above must be modified by exchanging $m$ and $m'$:
$m$ becomes
the strange mass, $m'$ the charm mass. The
GF contribution changes sign and so does the
QE
term. The difference of the whole
structure functions, $F_3^{\nu}
- F_3^{\bar \nu}$, is simply twice the $cs$ component for neutrino,
$2 \, F_3^{\nu, cs}$, since the
valence part cancels out (notice that
the $cs$ component disappears in the sum $F_3^{\nu}
+ F_3^{\bar \nu}$ and hence is irrelevant for
the Gross--Llewellyn Smith sum rule).

Let us now present the results of our calculations. The $\nu - \bar \nu$
difference of $xF_3$ structure functions,
$xF_3^{\nu} - x F_3^{\bar \nu}$, has been evaluated by using
eq.~(\ref{a7}) and the
MRS(A) fit \cite{MRS2} for the strange, charm
and gluon densities at the factorization scale $\mu^2$.
The choice of $\mu^2$ is a delicate issue. As we shall see, one of
the advantages of working with $xF_3^{cs}$ is that its dependence
on $\mu^2$ turns out to be quite small (smaller than the
$\mu^2$-dependence of $F_2^{cs}$). In any case, $\mu^2 = m_c^2$, where
$m_c$ is the charm mass, seems to
be a natural choice near (or not much above) threshold. At large $Q^2$
it is reasonable to take a factorization scale of order $Q^2$, rather
than $m_c^2$. The prescription of Ref.~\cite{Tung}, for instance,
gives $\mu^2 \simeq m_c^2$ at small $Q^2$ and $\mu^2 \simeq Q^2/2$
at $Q^2 \gg m_c^2$.

In Fig.~1 we present our results for $xF_3^{\nu}- xF_3^{\bar \nu}$
just above threshold ($Q^2 = 10$ GeV$^2$ and $25$ GeV$^2$),
with $\mu^2 = m_c^2$
(we use $m_c^2 = 2.7$ GeV$^2$). In this case $c(x,\mu^2)$ obviously vanishes.
The total result, eq.~(\ref{a7}), the quark excitation term,
the unsubtracted gluon fusion contribution, eq.~(\ref{a0}), and
the slow-rescaling prediction, eq.~(\ref{a8}),
are plotted in the figure.
An interesting feature is clearly visible: the complete NLO result
for $xF_3^{\nu}- xF_3^{\bar \nu}$ nearly coincides with the slow-rescaling
expectation, whereas it differs sensibly
from the unsubtracted GF result,
especially
for $x \gsim 0.01$.
Thus, the slow rescaling mechanism,
whose application to
the longitudinal+transverse structure function $F_2$ at small $Q^2$
is rather unsafe
(see \cite{BGNPZ5,BDG}),
turns out to be an excellent approximation when dealing with the
charm--strange contribution to $F_3$.
This is due to the fact that
the main next-to-leading order effects
that slow rescaling mimicks too  crudely
are
related to the longitudinal component of structure functions, which is
absent in $F_3$.
Notice also that the QE component, which is the only term containing the
strange density that experiments aim to
extract, is comparable in magnitude
to -- and actually not much different from -- the complete result.

We checked the dependence of the results on the factorization scale
$\mu^2$. This dependence is shown in Fig.~2 for two values of $x$  (0.01
and 0.1) and for $Q^2 = 25$ GeV$^2$ (close to the average $Q^2$ value of
the CCFR experiment). It is reassuring to see that, not only the total
result for $xF_3^{\nu}- xF_3^{\bar \nu}$, but each single term (and in
particular the QE term)
has a very mild dependence on $\mu^2$.
Moreover, the (small) $\mu^2$-dependence of the QE term is
approximately the same as that of the full $\nu - \bar \nu$
difference (the solid and dotted curves in the left part of Fig.~2
are parallel).
For comparison, in the right part of Fig.~2
we present also the scale dependence of $F_2^{\nu,cs}$, which appears
to be more dramatic, in particular in the QE component (for more details
we refer the reader to a forthcoming paper \cite{BDG}).
Since it is the latter component  that contains the strange and charm
densities, it is evident that the extraction of $xs(x)$ from
$xF_3^{\nu}- xF_3^{\bar \nu}$
 is  affected by a factorization scale uncertainty
much smaller than that occurring in a measurement based on
$F_2^{\nu,cs}$, such as the dimuon measurement.

At large $Q^2$ both strange and charm can be considered as massless
partons and one expects to regain the results of massless QCD: in particular,
the subtracted GF term should vanish. This is clearly visible in Fig.~3,
where
one can see that the full result coincides asymptotically both
with the QE term and with the
slow-rescaling prediction (this means that $F_{3,GF}^{\nu,cs}$ -- represented
by the dashed curve --
is exactly cancelled by the subtraction term, so that
$\tilde F_{3,GF}^{\nu,cs} =  0$).

Leaving aside the lack of statistics that may make the determination of
the strange density from $xF_3^{\nu} - x F_3^{\bar \nu}$
difficult in practice, it is clear
that this method has some indisputable advantages:
{\it i)} it is not plagued by relative-normalization errors; {\it ii)}
it is not affected by the ambiguities inherent in other methods, such as
the spurious separation of {\it t}- and {\it u}-channel diagrams that occurs in
the
dimuon separation (see \cite{BGNPZ4,BGNPZ5}); {\it iii)} large longitudinal
contributions arising from
the non-conservation of weak currents are obviously
absent; {\it iv)} charm mass effects are very well accounted for by
the slow-rescaling prescription, making the analysis of data and the
extraction of the strange density particularly simple (all the next-to-leading
order QCD machinery can be safely avoided); {\it v)} the dependence
on the factorization scale of the full $O(\alpha_s)$ result and of
the quark excitation
term, which contains the strange density to be extracted, is rather mild
and does not represent a worrisome source of uncertainty.

The main conclusion of our study is that
 a precision measurement of $xF_3^{\nu} - x F_3^{\bar \nu}$
may well represent
a new, interesting  source of information on the strange
quark distribution: being cleaner and more direct than other
determinations, it is certainly worth exploiting.

\vspace{1cm}

Useful discussions with A.~Bazarko and E.~Predazzi are gratefully
acknowledged.

\pagebreak

\pagebreak

\begin{center}
{\Large \bf Figure Captions}
\end{center}

\vspace{1cm}

\begin{itemize}

\item[Fig.~1]
The difference $xF_3^{\nu} - xF_3^{\bar \nu}$ at $Q^2 = 10$ GeV$^2$ and
25 GeV$^2$. The solid curve is the complete result, eq.~(\ref{a7}).
The dotted and dashed curves are the quark excitation (QE)
 and the
unsubtracted gluon fusion (GF)
contributions, respectively. The dot-dashed curve
is the slow-rescaling expectation, eq.~(\ref{a8}). The factorization scale
is $\mu^2 = m_c^2$.

\item[Fig.~2]
The dependence of $xF_3^{\nu} - xF_3^{\bar \nu}$ and of
$F_2^{\nu,cs}$ on the factorization scale $\mu^2$. The meaning of the curves
is the same as in Fig.~1.

\item[Fig.~3]
Same as Fig.~1 at $Q^2 = 100$ GeV$^2$ and 1000 GeV$^2$,
with $\mu^2 = Q^2/2$. Note that in the bottom window the solid, dotted and
dot-dashed curves coincide.

\end{itemize}

\end{document}